\documentclass[aps,a4paper,10pt,twocolumn,nofootinbib]{revtex4} 
\usepackage[T1]{fontenc}
\usepackage{mathptmx}
\usepackage{datetime}
\usepackage{amsmath}
\usepackage{amsfonts}
\usepackage{amsfonts}
\usepackage{mathrsfs}
\usepackage[mathscr]{euscript}
\usepackage[dvips]{graphicx}
\usepackage{fancyhdr}
\usepackage{colordvi}
\usepackage{hyperref} 
\usepackage{epsfig}
\usepackage{color}
\usepackage{bm}

\def\overstrike#1#2{{\setbox0\hbox{$#2$}\hbox to \wd0{\hss
    $#1$\hss}\kern-\wd0\box0}}

\renewcommand{\Vec}{\bm}

\newcommand{\indevice}[1]{\hat{#1}}
\newcommand{\indesign}[1]{\tilde{#1}}

\newdateformat{yymmdddate}{\THEYEAR/\twodigit{\THEMONTH}/\twodigit{\THEDAY}}

\def\FIGWIDTH{0.80}

\newcommand{\XDOI}[1]{\href{http://dx.doi.org/#1}{doi:#1}}
\newcommand{\XARXIV}[1]{\href{http://arxiv.org/abs/#1}{arXiv:#1}}

\begin{document}
\title{Cloak Imperfect: Impedance}
\author{Paul Kinsler}
\email{Dr.Paul.Kinsler@physics.org}
\affiliation{
  Physics Department,
  Lancaster University,
  Lancaster LA1 4YB,
  United Kingdom.}
\affiliation{
  Department of Physics, 
  Imperial College London,
  Prince Consort Road,
  London SW7 2AZ, 
  United Kingdom.
}

\lhead{\includegraphics[height=5mm,angle=0]{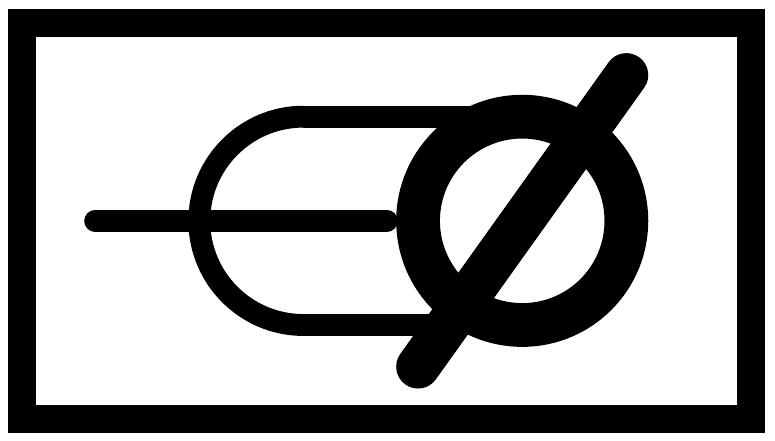}~~ZLOAK}
\chead{Cloaks/Impedances}
\rhead{
\href{mailto:Dr.Paul.Kinsler@physics.org}{Dr.Paul.Kinsler@physics.org}\\
\href{http://www.kinsler.org/physics/}{http://www.kinsler.org/physics/}
}

\begin{abstract}

I investigate the scattering properties of transformation devices
 as the traditional impedance matching criteria are altered.
This is demonstrated using simple theory
 and augmented by numerical simulations that investigate
 the role of impedance rescaling.
Results are presented for transformation devices
 in a cylindrical geometry, 
 but the lessons apply to both simpler
 and more complicated transformation devices.  
One technique used here is the use of impulsive field inputs, 
 so that scattered fields are more easily distinguished
 from non-scattered fields.

\end{abstract}


\date{\today}
\maketitle
\thispagestyle{fancy}

%

%
\section{Introduction}\label{S-intro}

Transformation Design --
 the use of the mathematical transformation of reference materials
 into those interesting ``device'' properties --
 is an area of active research interest.
Investigations range
 all the way from the most abstract theory and conceptualizing
 \cite{Kinsler-M-2015pnfa-tofu,Kinsler-M-2014adp-scast,Smolyaninov-2013jo,Mackay-L-2011prb,Thompson-CJ-2011jo}
 through to concrete theoretical proposals 
 \cite{Pendry-SS-2006sci,Lai-CZC-2009prl,Horsley-HMQ-2014sr,MitchellThomas-MQHH-2013prl,Kinsler-M-2014pra,McCall-FKB-2011jo,McCall-2013cp}
 and technological implementations
 \cite{Schurig-MJCPSS-2006s,Zhang-XF-2011prl,Frenzel-BBSKW-2013apl,Lukens-MLW-2014o}.

One aspiration is the ``perfect cloak'', 
 a Transformation Device (T-device) redirecting light, 
 sound, 
 or other signals 
 so that a fixed interior (core) region
 is invisible and undetectable by outside observers.
Although it seems that such a device is mathematically possible,
 there are many practical and technological constraints
 on what we can build that interfere with this ideal. 
Here I address the role of impedance rescaling,
 a common way of simplifying device designs.

As a start, 
 it was noted that the original radial cloaking conception
 was impedance matched at the boundary \cite{Pendry-SS-2006sci}.
However, 
 as far as standard impedance calculations are concerned,
 it was only impedance matched in the radial direction, 
 and not in the angular or axial directions --
 but it is worth also noting that 
 such naive uses of impedance measures in 
 the anisotropic materials generated by transformation design schemes
 give misleading results
 \cite{McCall-KT-2016jo-helimed,McCall-GK-2017perfract}.
In fact, 
 interfaces between any medium and a transformed version
 (e.g. that subject to a linear scaling perpendicular to the interface)
 are guaranteed to be reflectionless.

Nevertheless, 
 in the absence of a more general impedance calculation, 
 I will still use it as a benchmark about which 
 to consider impedance rescalings and their concomittant effect on 
 scattering from a set of transformation devices.
Here we will consider the simple example of a cylindrical transformation
 in a flat space, 
 as used in many cloaking designs.
The design is a purely spatial transformation
 applied to the EM constitutive parameters.
As described in \cite{Kinsler-M-2015raytail}
 such a transformation changes the effective metric as seen by 
 propagating electromagnetic fields.
But, 
 in changing the metric in order to ``steer'' the fields 
 as demanded by the transformation, 
 it does not specify anything about the impedance transformation.
It is left as a side effect of the constitutive transformation, 
 typically based on a ``kappa medium'' assumption 
 where $\kappa=\epsilon=\mu$
 (see e.g. \cite{McCall-KT-2016jo-helimed}),
 although other choices, 
 such as assuming a dielectric-only response, 
 are made depending on the situation or technological convenience.

\begin{figure}
\reflectbox{\includegraphics[width=\FIGWIDTH\columnwidth,angle=0]{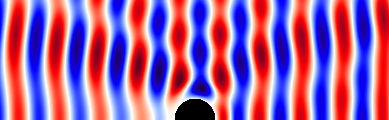}}\\
\vspace{1pt}
\reflectbox{\includegraphics[width=\FIGWIDTH\columnwidth,angle=0]{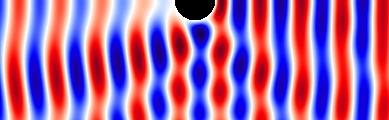}}
\caption{Finite difference time domain (FDTD)
 simulations of cloaking for plane wave (CW) sources
 on the left hand side.     
Results for two different cloaks, 
 as described later in Sec. \ref{S-simulations},
 are shown.
The upper plot shows 
 standard (linear) cloaking transformation,
 whereas
 the lower plot shows 
 a smoother logarithmic cloaking transformation.
Both show that the cloak functions as intended, 
 albeit imperfectly; 
 also that the differences between them are hard to discern by eye.
}
\label{fig-cloakingok}
\end{figure}

One notable feature of many reported cloaking results, 
 in either simulation or experiment,
 is that pictorial reprentations involve 
 the steady state situation with an incident plane wave 
 or other continuous wavefront.
These usually show, 
 as on fig. \ref{fig-cloakingok}, 
 a sufficiently convincing cloak performance, 
 albeit with the kind of imperfections one might expect --
 such as a slightly modulated or attenuated wave pattern, 
 providing evidence of scattering, 
 absorption, 
 or other imperfect implementation.
However it is rarely clear what specific feature of the model 
 gives rise to the imperfect performance.
Notably, 
 even numerical simulations are expected to have trouble near 
 the singular material properties
 at core of a cloak, 
 and we could well expect these to be 
 the dominant source of error.

In this paper 
 I investigate the role of impedance matching,
 as it is traditionally calculated, 
 and how altered impedance choices affect the overall scattering performance.
To do this I use
 snapshots from the impulsive probing of T-devices.
This use of an impulse 
 enables easy discrimination between scattering from the cloak halo
 and that from the core.
This shows that although the core provides the dominant failing, 
 the impedance matching also plays a role; 
 a distinction important if one imagines probing a cloak with a beam
 that misses the core.

%
\section{Linear Rescaling Transformation}\label{S-designeL}

It is worthwhile considering a simple linear rescaling
 as a T-Design.
Notably, 
 a one-axis rescaling turns out to be 
 a primitive building block for all T-devices.
We can see why this is by considering two nearby regions
 in a T-device, 
 which are infinitesimally different from one another 
 as the properties of the transformation change with position.
In order to maintain continuity of the transform over the contact surface
 between the two regions, 
 they can only differ by a rescaling perpendicular to that surface.
A two-axis linear rescaling cannot maintain continuity
 over a contact plane, 
 only over the line along the unscaled axis\footnote{Note that
   a gradual shear is also allowed, 
  but is not discussed here.}.

As a result, 
 at a sufficiently small scale,
 any transformation 
 (or \emph{morphism})
 is reducible to single axis anisotropic scaling,
 although the orientation of that rescaling axis 
 will typically vary with position.

Although this one axis rescaling has
 been treated previously \cite{McCall-KT-2016jo-helimed},
 I present an abbreviated version here.
If the $x$ and $y$ directions are chosen parallel 
 to the interface between two regions in a T-device, 
 whether these have a finite or infinitesimal extent, 
 we can rescale the $z$ axis in the second region by a factor $\lambda$.
Thus the $z$-direction refractive index squared 
 $n_z^2$
 must change by $\lambda^2$, 
 but the $x$ and $y$ direction counterparts ($n_x^2$ and $n_y^2$)
 remain unaltered.

Because EM is a transverse theory, 
 this means that the $x$ and $y$ material responses are scaled 
 by the transformations of the $z$ direction.
If the first medium was isotropic
 with $\epsilon=\mu=\kappa_1$, 
 the other is a new anisotropic ``$\kappa_\lambda$'' medium defined by 
~
\begin{align}
  \kappa_x = \epsilon_x = \mu_x 
&=
  \lambda \kappa_1,
\label{eqn-2-kappa-x}
\\
  \kappa_y = \epsilon_y = \mu_y
&= 
  \lambda \kappa_1,
\label{eqn-2-kappa-y}
\\
  \kappa_z = \epsilon_z = \mu_z 
&=
  \kappa_1
.
\label{eqn-2-kappa-z}
\end{align}

The impedances also change, 
 but of course there are two principal impedances
 per direction of propagation, 
 each being the reciprocal of the other.
The
 $z$ direction impedance squared is one of 
 $Z_z^2 = \left\{ \mu_x/\epsilon_y, \mu_y/\epsilon_x \right\}
 = 1$;
 the 
 $x$ direction impedance squared is one of 
 $Z_x^2 = \left\{ \mu_y/\epsilon_z, \mu_z/\epsilon_y \right\}
 = \left\{ \lambda, \lambda^{-1} \right\}$; 
 the 
 $y$ direction impedance squared is one of 
 $Z_y^2 = \left\{ \mu_x/\epsilon_z, \mu_z/\epsilon_x \right\}
 = \left\{ \lambda, \lambda^{-1} \right\}$.

Consequently, 
 {only} rays (or waves)
 that cross between the two regions 
 \emph{perpendicular} to the interface 
 see no calculated change in impedance;
 all others,
 to some extent,
 would be expected to probe
 the unmatched orientations.
Although this leads to a natural assumption that
 there will then be scattering or reflection from the interface,
 in fact
 these calculated impedance mismatches 
 do not result in reflections \cite{McCall-GK-2017perfract}.
This is a consequence of the fact that 
 at least for EM, 
 the usual transformation scheme preserves the solutions
 of Maxwell's equations at the same time as it redistributes
 the propagation.
This indicates that from a global perspective, 
 scattering from the transformation-induced calculated impedance changes
 need not occur\footnote{Note, 
 however, 
 that in a dynamical, 
 microscopic perspective, 
 the only boundary condition we are allowed to set is that 
 for the initial conditions.
Thus, 
 although some transformed solution
 may well \emph{still} be a solution of Maxwell's equations, 
 it may not necessarily be one accessed by dynamically evolving
 from specified boundary conditions.
However, 
 this possible loophole has subtle foundations, 
 and requires further (future) examination.
}, 
 even though from the traditional impedance perspective
 (e.g. \cite{Kinsler-2010pra-dblnlGpm})
 it clearly must.

\begin{figure}
  \begin{center}
    \resizebox{0.6\columnwidth}{!}{
    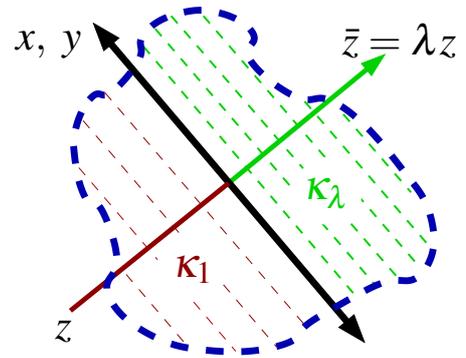}
  \end{center}
\caption[]
{ \label{fig-linsquash}
An interface between two (possibly infinitesimal) regions, 
 with a change in scaling between the ``ordinary'' medium 
 with properties given by $\kappa_1$,
 and the alternate medium which is scaled perpendicular to the interface
 by $\lambda$, 
 with properties given by $\kappa_\lambda$.
}
\end{figure}

%
\section{Radial Transformation Designs}\label{S-design}

Here we consider a 2D radial morphism
 where points with a laboratory or \emph{device} coordinate $\indevice{r}$
 are transformed so as to appear at some apparent or \emph{design} position 
 $\indesign{r}=f(\indevice{r})$, 
 just as in the T-Design for a cylindrical cloak.
Within this general approach,
 we can describe not only cloaks 
 but also various types of illusion and/or distortion devices,
 and even two universes connected
 by a wormhole \cite{Gratus-KM-2017nocharge}\footnote{To create such a
 ``biverse'' scenario by T-design, 
 we can set $\indesign{r}=1/\indevice{r}$ when $\indevice{r}<1$.}.
However, 
 we keep the mathematics general so that other non-cloak 
 morphisms are allowed by the theory presented here.
The original design for an electromagnetic cloak 
 \cite{Pendry-SS-2006sci}
 used a transformation based on simple linear scaling of the radius, 
 so here I will call that a linear-radial cloak.
Other forms are possible, 
 such as those based on polynomial forms (e.g. \cite{Cummer-LC-2009jap})
 or the natural logarithm 
 (see \cite{Cai-CKSM-2007apl,Kinsler-M-2015raytail} and later in this paper).

\begin{figure}
  \begin{center}
    \resizebox{\FIGWIDTH\columnwidth}{!}{
    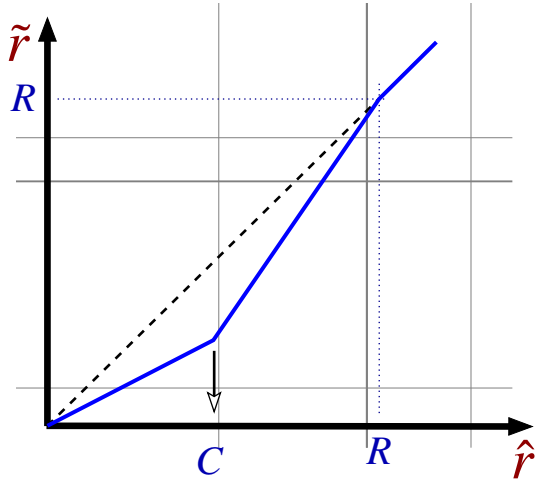}
  \end{center}
\caption{Radial distorting device, 
 with $\indevice{r}$ to $\indesign{r}=f(\indevice{r})$ mapping
 based on a piecewise linear scaling.
We could turn this into a standard linear cloak 
 by dropping $f(C)$ into the $\indevice{r}$-axis, 
 i.e. ensuring $f(C)=0$
}
\label{fig-radistort}
\end{figure}

\begin{figure}
  \begin{center}
    \resizebox{\FIGWIDTH\columnwidth}{!}{
    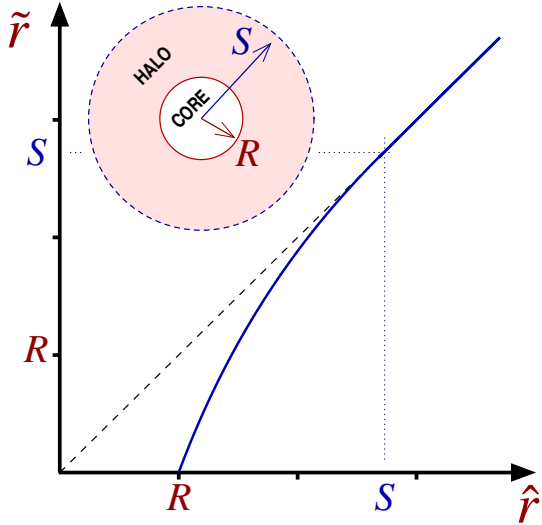}
  \end{center}
\caption{Radial cloak based 
 with an $\indevice{r}$ to $\indesign{r} = f(\indevice{r}) = R \log_{10}(e\indevice{r}/R)$ mapping 
 based on the logarithmic function
 (as in \cite{Ma-CYL-2009piers}).
The core boundary is at $\indevice{r}=R/e$, 
 and the halo boundary -- 
 its interface with the exterior --
 is at $\indevice{r}=R$.
}
\label{fig-radcloak}
\end{figure}

Here I will primarily consider three devices:
 (a) a radial distorter based on a piecewise linear transformation 
 where objects in the core region
 will appear to an outside observer 
 to have a smaller size
 (see fig. \ref{fig-radistort}), 
 (b) a smoothed radial distorter based on a cosine transformation, 
 and
 (c) a smooth radial cloak based on a logarithmic transformation
 (see fig. \ref{fig-radcloak}).

Partly following \cite{Cummer-LC-2009jap}, 
 we define a radial transformation from device radial coordinate $\indevice{r}$
 to design (apparent) coordinate $\indesign{r}$:
~
\begin{align}
  \indesign{r} &= f(\indevice{r})
\\
  \textrm{with} \qquad
  f'(\indevice{r}) 
&=
  \frac{\partial f(\indevice{r})}{\partial \indevice{r}} 
=
  \frac{\partial \indesign{r}}{\partial \indevice{r}}
\end{align}

This means that 
~
\begin{align}
  \kappa_r = \epsilon_r = \mu_r 
&=
  \frac{f(\indevice{r})}{\indevice{r} f'(\indevice{r})}
,
\label{eqn-fr-kappa-r}
\\
  \kappa_\theta = \epsilon_\theta = \mu_\theta 
&= 
  \frac{\indevice{r} f'(\indevice{r})}{f(\indevice{r})}
,
\label{eqn-fr-kappa-theta}
\\
  \kappa_z = \epsilon_z = \mu_z 
&=
  \frac{f'(\indevice{r}) f(\indevice{r})}{\indevice{r}}
.
\label{eqn-fr-kappa-z}
\end{align}

For a cloak, 
 the finite radius $\indevice{r}=C$ of the core region (in the laboratory)
 needs to behave as if contracted to the origin $\indesign{r} = f(\indevice{r}) = 0$.
This means that no matter what $f(\indevice{r})$ we define, 
 one of $\kappa_r$ or $\kappa_\theta$ will diverge near $\indevice{r}=C$.
Most likely this will be $\kappa_\theta$, 
 since typically $f'(\indevice{r})$ will be finite, 
 although if we engineered $f'$ to vanish faster than $f(\indevice{r})$ 
 then $\kappa_r$ would diverge.
The intermediate situation where $f' \propto f$, 
 which would allow non-singular properties,
 also means that $f$ is an exponential function, 
 which has the wrong behaviour to be used for a cloak\footnote{It
  would be an anti-cloak, 
  where a visibly missing disk were represented
  in the device by all points down to $\indevice{r}=0$; 
  although with a minus sign we could instead allow all $\indevice{r}$
  to represent a disk.}.

%
\subsection{Index $n$}\label{S-cylinder-index}

Given a T-Design
 defined by the function $\indesign{r}=f(\indevice{r})$, 
 we can directly find 
 the material refractive indexes
 it needs to work.
Assuming the design is to look like a space
 with fixed index normalized to $1$, 
 they are 
~
\begin{align}
  n_r^2 
&= 
  \left\{ \epsilon_\theta \mu_z , \epsilon_z \mu_\theta \right\}
  =
    \kappa_\theta \kappa_z
  =
    f'(\indevice{r})^2
\\
  n_\theta^2
&=
  \left\{ \epsilon_z \mu_r , \epsilon_r \mu_z \right\}
  =
    \kappa_r\kappa_z 
  =
    \frac{f(\indevice{r})^2}{\indevice{r}^2}
\\
  n_z^2
&=
  \left\{ \epsilon_r \mu_\theta , \epsilon_\theta \mu_r \right\}
  =
    \kappa_r \kappa_\theta
  =
    1
\end{align}
The second term on each line indicates that there are
 two ways of making up each index from the 
 underlying constitutive parameters 
 (i.e. the permittivity and permeability).
Typically we take this as an opportunity to restrict our design
 to only one of these polarizations, 
 but for a \emph{perfect} cloak both would have to be allowed for.

We see here that even 
 for a cloak,
 it is trivial to ensure the index profiles are non singular, 
 although 
 $n_\theta$ does vanish on the inside core edge.
It also looks relatively simple to index-match 
 $n_r$ at the outer boundary by a suitable choice of gradient $f'$, 
 if there were a reason to do so.
Note that 
 the axial index $n_z$ is always the 
 same as that of the background index.

%
\subsection{Impedance $Z$}\label{S-cylinder-impedance}

Now, 
 based on the ``standard'' changes in $\epsilon$ and $\mu$ 
 as specified in eqns. \eqref{eqn-fr-kappa-r}, 
  \eqref{eqn-fr-kappa-theta}, \eqref{eqn-fr-kappa-z}, 
 we can calculate 
 the impedances in the way they are usually defined.
Each direction of propagation has an impedance that 
 depends on the field polarization, 
 and is derived from two principal values for that direction.

The radial impedance $Z_r$
 has principal values which are
~
\begin{align}
  Z_r^{2} 
&=
  \left\{ \frac{\mu_z}{\epsilon_\theta} , \frac{\mu_\theta}{\epsilon_z} \right\}
=
  \left\{
    \frac{f(\indevice{r})^2}{\indevice{r}^2}, \frac{\indevice{r}^2}{f(\indevice{r})^2}
  \right\}
.
\end{align}
The radial impedance is therefore matched at the outer (halo) boundary
 where $R=f(R)$; 
 note that this is true for any cloak, 
 not just linear ones \cite{Pendry-SS-2006sci}.

The angular impedance $Z_\theta$
 has principal values which are
~
\begin{align}
  Z_\theta^{2} 
&=
  \left\{ \frac{\mu_r}{\epsilon_z} , \frac{\mu_z}{\epsilon_r} \right\}
=
  \left\{
    f'(\indevice{r})^{-2} , f'(\indevice{r})^2
  \right\}
.
\end{align}
{Unlike} the radial impedance, 
 the angular impedance is \emph{not} matched at the outer boundary
 unless our design is such that $f'(R)=1$.
Thus it is unmatched for the linear cloak --
 which was therefore \emph{not} perfectly impedance matched, 
 but is true for (e.g.) the logarithmic cloak.

The axial impedance $Z_z$
 has principal values which are
~
\begin{align}
  Z_z^{2} 
&=
  \left\{ \frac{\mu_\theta}{\epsilon_r} , \frac{\mu_r}{\epsilon_\theta} \right\}
=
  \left\{
    \frac{r^2 f'(\indevice{r})^2}{f(\indevice{r})^2} , \frac{f(\indevice{r})^2}{\indevice{r}^2 f'(\indevice{r})^2} 
  \right\}
.
\end{align}
As for the radial impedance, 
 the linear cloak again fails impedance matching, 
 but if we have a design where $f'(R)=1$
 then it will be impedance matched at the outer boundary.

For a cloak, 
 where $f(\indevice{r})=0$ when $\indevice{r} \neq 0$ 
 at the core boundary,
 both $Z_r$ and $Z_z$ are zero or singular, 
 depending on polarization.
At the halo boundary 
 $Z_r$ is unity (background), 
 since $R=f(R)$,
 while the others depend on the gradient $f'$.
As long as the design ensures that $f'(\indevice{r})$ stays non-zero, 
 the angular impedance need never be singular.

We might try to  reduce the singularities in material properties
 at the core boundary by means of an $f(\indevice{r})$ that skims 
 at a vanishingly low angle into the $r$ axis, 
 so that $f'(\indevice{r}) \rightarrow 0$
 (see e.g. \cite{Cummer-LC-2009jap}).
Although this would be expected to reduce scattering 
 by reducing the requirement for impractical material properties, 
 this gain could well be balanced by an increase 
 in impedance-derived scattering.

%
\subsection{Transform $Z$ while preserving $n$}\label{S-cylinder-transform}

We now consider rescaling impedances by multiplying 
 all permittivity values by a factor $\xi$
 whilst dividing
 all permeability values by that same factor. 
The refractive indexes will then remain constant, 
 preserving the cloak's ``steering'' 
 properties, 
 but its impedance matching is altered.

For an electromagnetic cloak, 
 we can choose to look mainly at the $r,\theta$ plane, 
 and consider electric fields
 aligned only in the $r,z$ plane\footnote{Choosing
  magnetic fields in $r,z$ instead
  produces complementary results, 
  with an otherwise almost identical character --
  it only means that the relevant $\bar{Z}_z$ impedance
 is the other (reciprocal) choice
 out of the two possible principal values.}, 
 so that only $\epsilon_z$, $\mu_\theta$, and $\mu_r$ are relevant.
If we choose the factor $\xi=\indevice{r}/f(\indevice{r})$,
 then we can cancel the radial impedance profile
 inside the cloak.
This makes the radial impedance have 
 the same constant value inside the cloak as it has outside,
 and even a traditional interpretation would hold that 
 no radially-travelling components would be reflected (scattered).
However, 
 this is at the cost of altering the variation
 in the angular profile $\bar{Z}_\theta^{2}$
 which now depends on $f(\indevice{r})$.
Unfortnately, 
 $f(\indevice{r})$ causes problems, 
 because it vanishes at $\indevice{r}=C$.
We find that 
~
\begin{align}
  \bar{Z}_r^{2}
&=
  \frac{\bar{\mu}_\theta}{\bar{\epsilon}_z}
   = \frac{{\mu}_\theta/\xi}{{\xi\epsilon}_z}
   = \left[ f' \right]^{-1} . {f'} 
   = 1
,
\\
  \bar{Z}_\theta^{2}
&=
   \frac{\bar{\mu}_r}{\bar{\epsilon}_z}
   = \frac{{\mu}_r/\xi}{\xi{\epsilon}_z}
   = \left[ \frac{f^2}{\indevice{r}^2 f'} \right] .  {f'}^{-1}
   = \frac{f^2}{\indevice{r}^2 f'^2}
,
\\
  \bar{Z}_z^{2}
&=
   \frac{\bar{\mu}_\theta}{\bar{\epsilon}_r}
   = \frac{{\mu}_\theta/\xi}{\xi{\epsilon}_r}
   = \left[ f' \right] . {f'}
   = {{f'}^2}
.
\end{align}
Here, 
 for a cloak, 
 none of the rescaled constitutive parameters need diverge
 near the core boundary,
 although $\bar{\mu}_r$ does tend to zero there.

Alternatively,
 by choosing $\xi=1/f'(\indevice{r})$ 
 we can fix the angular impedance inside the cloak 
 to be the same as that outside. 
However, 
 the radial impedance is no longer matched at the halo boundary
 unless $f'(b)=1$.
We find that 
~
\begin{align}
  \bar{Z}_r^{2}
&=
  \frac{\bar{\mu}_\theta}{\bar{\epsilon}_z}
   = \frac{{\mu}_\theta/\xi}{\xi{\epsilon}_z}
   = \left[ \frac{\indevice{r} {f'}^2 }{f} \right]^{-1} . \frac{f}{\indevice{r}} 
   = \frac{\indevice{r}^2 {f'}^2}{ f^2}
,
\\
  \bar{Z}_\theta^{2}
&=
  \frac{\bar{\mu}_r}{\bar{\epsilon}_z}
   = \frac{{\mu}_r/\xi}{\xi{\epsilon}_z}
   = \left[ \frac{f}{\indevice{r}} \right] . \left(\frac{f}{\indevice{r}}\right)^{-1}
   = 1
,
\\
  \bar{Z}_z^{2}
&=
  \frac{\bar{\mu}_\theta}{\bar{\epsilon}_r}
   = \frac{{\mu}_\theta/\xi}{\xi{\epsilon}_r}
   = \left[ \frac{\indevice{r} {f'}^2 }{f} \right] . \left( \frac{f}{\indevice{r}{f'}^2} \right)^{-1}
   = \frac{\indevice{r}^2 {f'}^4}{f^2}
.
\end{align}
Here,
 for a cloak, 
 we see that the $\bar{Z}_r$ and $\bar{Z}_z$ impedances
 are singular at the core boundary
 as a consequence of the rescaled constitutive parameters
 diverging or becoming zero.

Lastly, 
 we might decide to fix the $z$ impedance inside the cloak 
 to that outside.
That is, 
 we scale using $\xi=\indevice{r}f'/f$, 
 so we find that 
~
\begin{align}
  \bar{Z}_r^{2}
&=
  \frac{\bar{\mu}_\theta}{\bar{\epsilon}_z}
   = \frac{{\mu}_\theta/\xi}{\xi{\epsilon}_z}
   =  \left[ 1 \right] . \left( \frac{{f'}^2}{\indevice{r}^2} \right)^{-1}
   = \frac{\indevice{r}^2}{{f'}^2}
\\
  \bar{Z}_\theta^{2}
&=
  \frac{\bar{\mu}_r}{\bar{\epsilon}_z}
   = \frac{{\mu}_r/\xi}{\xi{\epsilon}_z}
   = \left[ \frac{f^2}{\indevice{r}^2{f'}^2} \right] . \left( \frac{f^2}{\indevice{r}^2} \right)^{-1}
   = {f'}^{-2}
\\
  \bar{Z}_z^{2}
&=
  \frac{\bar{\mu}_\theta}{\bar{\epsilon}_r}
   = \frac{{\mu}_\theta/\xi}{\xi{\epsilon}_r}
   = \left[ 1 \right] . 1
   = 1
.
\end{align}

From this
 there are three distinct impedance rescalings that seem useful.
We might continuously tune between the standard $\xi=1$ case
 with three parameters $a, b, c$,
 each adding in some degree those three cases
 discussed above.
The combined scaling parameter
 is then $\xi = \xi_r \xi_\theta \xi_z$, 
 where

\begin{enumerate}
%

\item
$a$ scales towards the 
perfect radial $Z_r=1$ case
 by setting $\xi_r=(\indevice{r}/f)^a$.

%

\item
$b$ scales towards the 
perfect angular $Z_\theta=1$ case 
 by setting $\xi_\theta= (1/f')^b$.

\item
$c$ scales towards the 
perfect axial $Z_z=1$ case 
 by setting $\xi_z= (\indevice{r}f'/f)^c$.

\end{enumerate}

From the above, 
 we can see that the rescaling that fixes the radial impedance
 seems the best behaved:
 assuming $f'$ is well behaved, 
 the sole remaining difficulty is with the vanishing value of $\mu_r$
 at the inner boundary.
Also, 
 any propagation along a predominantly angular path
 will be a mostly constant radius, 
 and such propagation will not see any variation in impedances,  
 which change only with radius.

%
\section{Results}\label{S-simulations}

Although a perfectly implemented T-device should 
 exhibit no unwanted scattering or reflections,
 we expect that one whose impedances have 
 been rescaled as discussed in the preceeding section
 will do so.
In such a case the traditional perspective, 
 where impedance changes generate reflections and/or scattering
 again becomes relevant,
 and ensuring that there is no step-change 
 along one given direction
 might have a trade-off involving 
 the matching along other directions.
Further, 
 while asking for continuity in impedance
 is (\emph{was}) usually the most important demand, 
 impedance \emph{gradients} would also be considered causes of reflections, 
 albeit distributed ones that are likely to be weak.

To make these ideas about the effects of impedance matches, 
 mismatches, 
 and gradients more concrete I have performed sets of numerical simulations
 using MEEP \cite{Oskooi-RIBJJ-2010cpc}
 for various impedance criteria.
These were done in 2D (i.e. the $x, y$ plane), 
 and for transverse electric fields $\Vec{E}$, 
 so that $E_z \neq 0$ but $H_z=0$.
Further, 
 although cloaking devices are typically more interesting than
 simple distorters of the type shown in fig. \ref{fig-radistort}, 
 they have singular material properties at their core boundary.
Such singular properties give rise to numerical difficulties, 
 and ones that potentially will obscure the impedance-based properties
 of interest here.
Consequently I mainly show results for distorting devices,
 where it is easier to guarantee that the material properties 
 are well behaved.

I consider two sample distorting functions, 
 where $f_i(\indevice{r}) = \indevice{r} + h_i(\indevice{r})$.
Both depend on a parameter $\alpha$, 
 which specifies the (same)
 maximum displacement in either case.
Further, 
 it occurs at the same point $\indevice{r}=R/2$
 so that $h_i(R/2) = - \alpha R/2$.
The functions $h_i(\indevice{r})$ are

\begin{enumerate}

\item
A piecewise linear distortion, 
 where
~
\begin{align}
\qquad
  h_1(\indevice{r}) &= - \alpha \indevice{r}; 
&\textrm{for} \quad 0 < \indevice{r} \le R/2
\\
\qquad
  h_1(\indevice{r}) &=  \alpha \left( \indevice{r} - R/2 \right); 
&\textrm{for} \quad R/2 \ge \indevice{r} < R
.
\end{align}
This is continuous, 
 but does not match gradients.
To ensure that $f(r)$ is single valued and always increasing, 
 it requires $|\alpha| < 1$.

\item
A smoothly varying distortion, 
 with 
~
\begin{align}
  h_2(\indevice{r}) 
&=
  \frac{\alpha R}{4}
  \left[
    \cos \left( \frac{2\pi \indevice{r}}{R} \right) - 1
  \right]
.
\end{align}
This is both continuous
 and matches the gradient near the origin and at the boundary.
Since
\begin{align}
  h_2'(\indevice{r}) 
&=
  \frac{\alpha \pi}{2}
    \sin \left( \frac{2\pi \indevice{r}}{R} \right)
,
\end{align}
 we require that $\alpha < 2/\pi$
 to ensure that $f(\indevice{r})$ is single valued and always increasing.

\end{enumerate}

\begin{figure}
  \fbox{\includegraphics[width=0.3\columnwidth,angle=-0]{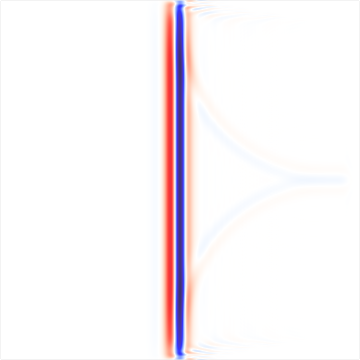}}
  \fbox{\includegraphics[width=0.3\columnwidth,angle=-0]{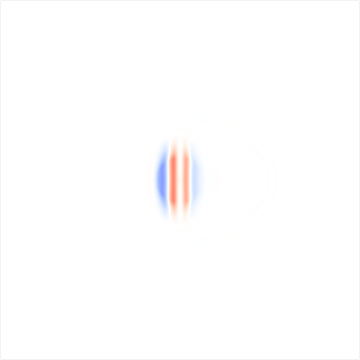}}
  \fbox{\includegraphics[width=0.3\columnwidth,angle=-0]{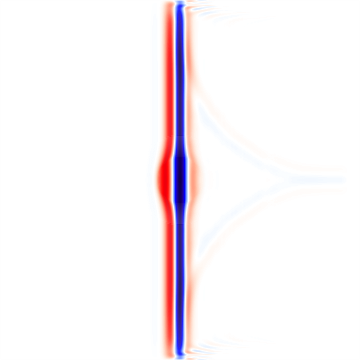}}
\caption{
Comparison of electric field profiles $E_z$
 taken when a planar wave packet, 
 travelling from right to left,
 is half way across a piecewise linear distorting T-device
 with $\alpha=1/3$.
To the left and right
 are the reference and distorting cases respectively; 
 in the centre
 the difference between them, 
 which also reveals the location of the T-device.
The faint curved wavefronts trailing the main wave 
 are artifacts of the source setup, 
 but are shared by both reference and T-device simulations.
}
\label{fig-distort-compare}
\end{figure}

\begin{figure}
  \includegraphics[width=0.484\columnwidth,angle=-0]{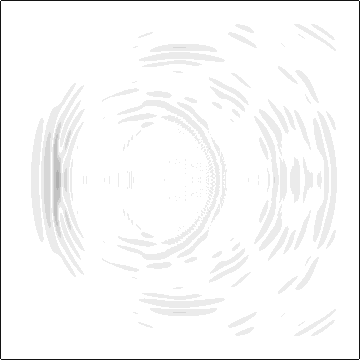}
  \includegraphics[width=0.484\columnwidth,angle=-0]{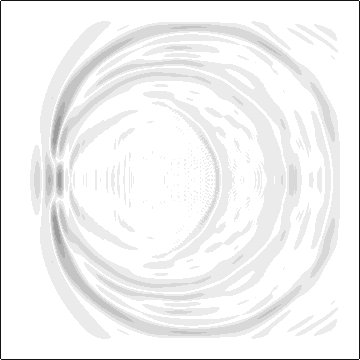}
  \includegraphics[width=0.484\columnwidth,angle=-0]{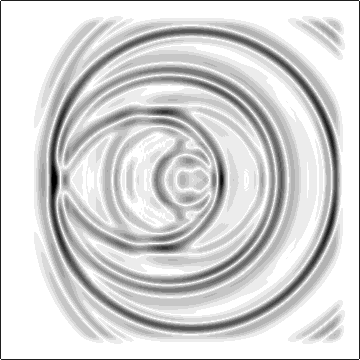}
  \includegraphics[width=0.484\columnwidth,angle=-0]{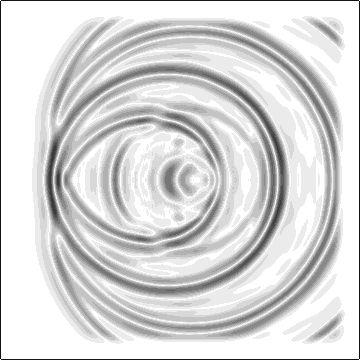}
\caption{
 Net scattering $E_z-E_{0}$ from a piecewise linear
 non-cloaking transformation.
A planar wave packet enters from the left hand side and traverses the 
 distorting region in the centre.
Because these plots only show the difference from the reference case, 
 the wave packet is almost entirely cancelled out.
However, 
 if it were visible, 
 it would form a vertical bar on the left hand side of each frame shown, 
 coincident with the left-most scattered fields.
The upper left frame shows the standard case,
 with a small amount of scattering; 
 but since theoretically this design should be a perfect cloak, 
 we can take this discrepancy to be an indication of 
 the numerical error.
The upper right frame
 shows 
 the 
  fixed matched radial impedance case.
The lower frames show 
 fixed matched angular impedance (left),
 and
 fixed matched axial impedance (right).
The contrast ratio has been increased by a factor of $20$ 
 over that for an ordinary non-difference plot.
}
\label{fig-distort-scatterings}
\end{figure}

The main quantity of interest here is the difference between 
 a numerical simulation based on a T-device
 and another simulation based on the unremarkable design space
 the T-device is intended to mimic.
In fig. \ref{fig-distort-compare} we can see such a comparison, 
 depicted as a wave packet crossed the centre of a distorting T-device.
However, 
 the actual differences of interest are ones 
 evaluated at a time after the input wave has 
 passed through and then left the T-device, 
 as well as 
 after the bulk of the scattered waves have also departed it; 
 but not before they reach the absorbing boundaries of the simulation edges.

However, 
 before proceeding with more exhaustive comparisons, 
 consider the scattering from the piecewise linear transformation above
 for different choices of impedance matching.
This is shown pictorially in fig. \ref{fig-distort-scatterings}
 for four different cases, 
 where the difference between the distorting simulation 
 and a reference simulation with a homogeneous background material
 is taken.
Since we choose a TE polarization for the simulations, 
 the difference in the $z$ (axial) component of the electric fields
 is plotted.
As discussed above, 
 in each case the index profile of the material is the same, 
 but different impedance criteria are imposed.
Clearly
 the different cases give rise to different scattering levels.

In what follows 
 we will combine the selected data from each specific simulation 
 as shown pictorially in fig. \ref{fig-distort-scatterings}
 into a single numerical value
 representing the scattering induced by the distorting T-device.
Each summed scattering value $S$ is calculated 
 from the field values
 $E_z^{(i)}(x,y)$ and $E_z^{(0)}(x,y)$
 for the distorting simulation
 and the reference simulation respectively.
The calculation is 
~
\begin{align}
  S^{(i)}
&=
  \int
    \int  
      \left[
        E_z^{(i)}(x,y) - E_z^{(0)}(x,y)
      \right]^2
   dx ~
 dy
,
\end{align} 
 although below we will plot $\log_{10} (S)$
 to enhance the level of detail visible on the figures.
Note also that the $S$ values are unnormalised sums over the numerical data,
 and not corrected for (e.g.) simulation resolution.

\begin{figure}
  \includegraphics[width=\FIGWIDTH\columnwidth,angle=-0]{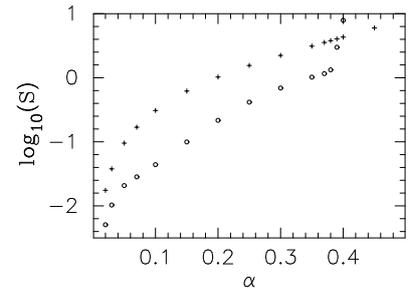}
\caption{
 Net numerical scattering vs distortion strength $\alpha$
 for both linear ($+$) and smooth cosine ($\circ$)
 non-cloaking transformation, 
 shown using a logarithmic scale.
This is for the standard impedance $\kappa$ medium choice
of $A=B=0$ .
}
\label{fig-distort-alpha}
\end{figure}

In fig. \ref{fig-distort-alpha} we can see how scattering increases
 for the standard impedance choice
 as the level of distortion is increased.
If you take the position that in-principle transformation devices 
 are capable of being perfect,
 as indicated by the lack of a reflection
 from a transformation-derived interface \cite{McCall-GK-2017perfract},
 then this figure provides a benchmark for the 
 numerical error in the simulations.

The smooth cosine distortion usually gives less scattering, 
 except as $\alpha$ approaches $1/2$, 
 when its transform generates regions of extreme stretching
 (where $f'(r) \rightarrow 0$).
Although the piecewise linear distortion has the disadvantage
 of abrupt interfaces, 
 the cosine distortion has regions that are more stretched, 
 which can override the benefits of smoothness.

The next step is a more thorough search of the impedance rescaling 
 parameter space for the two types of distorting T-device considered here.
In the previous section, 
 we said that if we take each of the ``obvious'' scalings in turn, 
 each raised to some power $a$, $b$, and $c$, 
 then the scaling factors will be  
 $(\indevice{r}/f)^a$, $(1/f')^b$, or $(\indevice{r} f'/f)^c$.
The net scaling in such a case is then 
 $(\indevice{r}/f)^A f'^{A-B}$ with $A=a+c$ and $B=b+a$.
This means that instead of displaying a 3D dataset 
 over the range of interesting $a,b,c$ 
 it is sufficiently instructive to check just the 
 2D range $A,B$.
Note that if $A=B$ then the scaling is $(\indevice{r}/f)^A$, 
 so that if $A=B=1$
 then we have fixed the radial impedance at a fixed value which is
 impedance matched to the background space.
If instead we choose $A=0$,
 then the scaling is $(1/f')^B$, 
 so that if $B=1$ we have fixed the angular impedance at a value 
 impedance matched to the background space.
Lastly, 
 if $B=0$ then the scaling is $(\indevice{r}f'/f)^A$, 
  so that at $A=1$ we have fixed the axial ($z$) impedance at a value 
 impedance matched to the background space.

\begin{figure}
  \includegraphics[width=\FIGWIDTH\columnwidth,angle=-90]{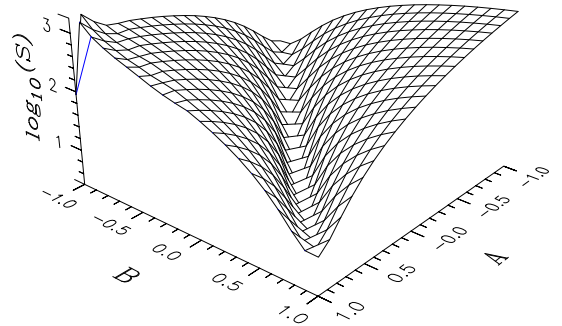}
\caption{
 Net scattering $S$ vs impedance rescaling
 parameters $A, ~B$
 for a piecewise linear
 non-cloaking transformation, 
 shown using a logarithmic scale.
These results were obtained 
 for a distortion strength of $\alpha=1/3$.
}
\label{fig-distort-linear}
\end{figure}

\begin{figure}
  \includegraphics[width=\FIGWIDTH\columnwidth,angle=-90]{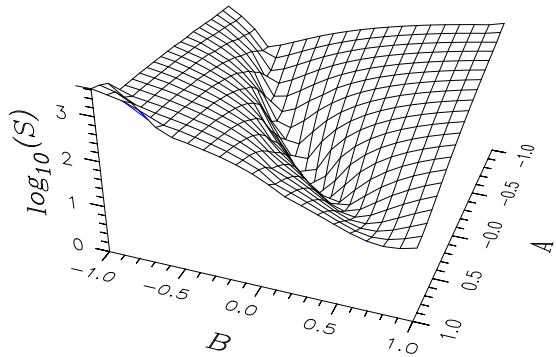}
\caption{
 Net scattering $S$ vs impedance rescaling
 parameters $A, ~B$
 for the smooth cosine
 non-cloaking transformation, 
 shown using a logarithmic scale.
These results were obtained 
 for a distortion strength of $\alpha=1/3$.
}
\label{fig-distort-cosine}
\end{figure}

In figs. \ref{fig-distort-linear} 
 and \ref{fig-distort-cosine}
 we see the excess scattering of 
 the two distorting T-devices.
In both cases the best performance is at the standard case
 where $A=B=0$, 
 with a slight degradation in performance 
 away from the origin along the line $A=B$; 
 and a strong degradation along $A=-B$.

\begin{figure}
  \includegraphics[width=\FIGWIDTH\columnwidth,angle=-90]{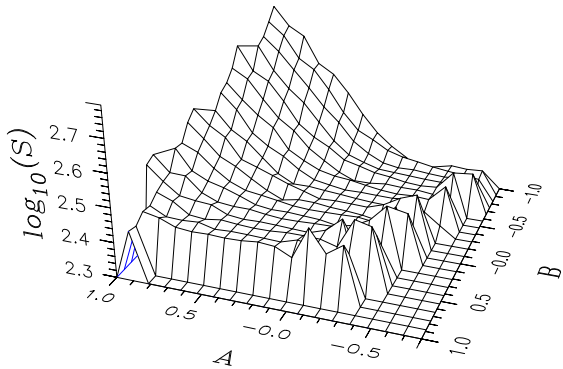}
\caption{
 Net scattering $S$ vs impedance rescaling
 for the log-based cloaking transformation, 
 shown using a logarithmic scale.
In cases where
 the extreme material properties caused numerical difficulties,
 the $\log_{10}(S)$ values were set to the 
 convenient value of +2.3 to aid presentation of the results.
}
\label{fig-SCloakCf-log} 
\end{figure}

\begin{figure}
  \includegraphics[width=\FIGWIDTH\columnwidth,angle=-90]{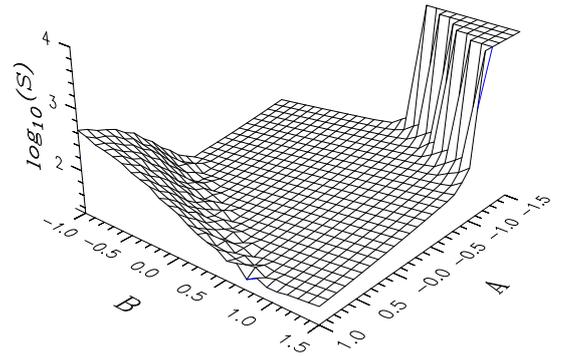}
\caption{
 Net scattering $S$ vs impedance rescaling
 for the log-based cloaking transformation acting to shrink a
 metallic scatter, 
 shown using a logarithmic scale.
In the corner characterised by large negative $A$ and 
 large positive $B$, 
 the tendency for increased scattering
 is quickly overwhelmed by parameter ranges where  
 the simulations become unreliable due to extreme material properties.
Here the $\log_{10}(S)$ values were set to the 
 convenient value of +4 in such cases.
}
\label{fig-SCloakCg-log} 
\end{figure}

It is also possible 
 to do similar comparisons of cloaking T-devices
 rather that the distorting ones presented here.
However, 
 for both the linear cloaking transformation
 and a smoother logarithmic transformation, 
 the scattering was dominated by the singular behaviour
 at the core boundary.
Further, 
 as can be seen in fig. \ref{fig-SCloakCf-log}, 
 the impedance rescaling exacerbated numerical difficulties in some cases, 
 so that a smaller range of rescalings gave useful results.

Therefore, 
 in order to enable investigation of a wide parameter space, 
 the cloak core was replaced with a larger metallic scatterer.
The cloak transformation then acted simply to shrink the effective size
 of this scatterer.
The comparison, 
 then, 
 is between the simulation of the cloak-based shrunk scatterer
 and a reference simulation with a scatter of the smaller (shrunken) size.
For this case, 
 only for larger values of $|A+B|$
 could the impedance-induced scattering be seen 
 over the other differences.
See, 
 for example, 
 fig. \ref{fig-SCloakCg-log}, 
 where the excess scattering is shown 
 for a logarithmic cloaking function
 where $f(\indevice{r}) = R \log (e\indevice{r}/R)$.
Unlike the narrow valley features seen in 
 fig. \ref{fig-distort-linear} and \ref{fig-distort-cosine}, 
 the central part of the parameter space 
 consists of a broad plateau.
Note that 
 due to the different simulation parameters, 
 these cloaking-based $S$ values are not directly comparable 
 to the distortion-based ones.

%
\section{Summary}\label{S-summary}

Here we have seen 
 that the usual $\epsilon=\mu$ transformation medium 
 provides the best performance in numerical simulations, 
 with a minimum of extraneous reflections and scattering 
 from the boundary and interior of the transformed region.
Although  this result was to be expected, 
 since on theoretical grounds the scattering should be exactly zero, 
 a traditional optics view of impedance matching
 would not necessarily have supported such a conclusion.
This is because, 
 as shown here, 
 a radial transformation such as that
 of the original Pendry et al. cylindrical cloak only matches --
 in the traditional sense --
 the radial impedance,
 with the angular and axial impedances left unmatched.
Attempts to improve impedance matching 
 (and so reduce scattering)
 by complementary rescalings of $\epsilon$ and $\mu$
 were unsucessful; 
 although not every possible rescaling was tested.
It is therefore clear that the meaning of impedance 
 is not well defined in the rather general types of anisotropic media
 that result from T-design.  

One further conclusion that we can draw, 
 is that it is important to be cautious
 when trying to improve on cloaking design, 
 as in the scheme of Cummer et al. \cite{Cummer-LC-2009jap}.
Although such re-designs may remove or moderate singularities
 in material parameters, 
 unless we can build the perfect $\epsilon=\mu$ device,
 the re-design may at the same time exacerbate impedance mismatches, 
 leading to a scattering increase instead of the intended decrease.
The use in this paper of an impulse wave profile to probe
 T-device performance was important in generating and understanding 
 the results presented here -- 
 the scattered wave can be directly seen in pictorial plots,
 as well as after summation of the net scattering.


%
\acknowledgments
I acknowledge valuable discussions with Martin McCall,
 Robert Thompson, 
 and Jonathan Gratus.
The great majority of the work here was done whilst 
 at Imperial College London, 
 supported by EPSRC (grant number EP/K003305/1); 
 but the final updates when at Lancaster University, 
 again supported by EPSRC (the Alpha-X project EP/N028694/1).

%


\end{document}